\documentstyle[amssymb,aps,twocolumn,floats,psfig,prb,tighten]{revtex}

\begin{document}

\draft 

\title{CPA for charge ordering in the extended Hubbard model}
\author{ A. T. Hoang$^{1,2}$ and P. Thalmeier$^3$}
\address{ 
$^1$\ Max-Planck-Institut f{\"u}r Physik komplexer Systeme,
         N{\"o}thnitzer Strasse 38, 01187\\
         Dresden, Germany}
\address{
$^2$\ Institute of Physics, P.O.Box 429 Bo Ho, Hanoi 10000, Vietnam}
\address{
$^3$\ Max-Planck-Institut f{\"u}r Chemische Physik fester Stoffe,
       N{\"o}thnitzer Strasse 40, 01187 Dresden, Germany}

\date{\today}
\maketitle

\begin{abstract}
We study charge ordering in the extended Hubbard model with both on-site 
and nearest neighbor Coulomb repulsion ($U$ and $V$, respectively) within 
the Coherent potential approximation (CPA). The phase boundary between the 
homogeneous and charge ordered phase for the square lattice is obtained 
for different values of $U$. It is shown that at quarter filling for 
all values of $U$ the charge ordering exists only if the inter-site 
Coulomb repulsion $V$ exceeds certain critical value  which is of the 
order of the kinetic energy $t$. At finite temperature a reentrant 
transition is found in some region of $V$.
\end{abstract}

\vspace{0.5cm}

\section {Introduction}  

 The problem of charge ordering (CO) attracted the attention of physicists 
already at the end of the thirties. In the low density limit, as was 
first proposed by Wigner \cite{b.a1}, the electrons crystallize in form 
of a lattice 
in order to keep the Coulomb repulsion as small as possible. Such a 
Wigner lattice is experimentally realized in a GaAs/AlGaAs 
heterostructure 
\cite{b.a2}. The CO may also occur at higher electron concentration if the 
interaction of electrons with spin degrees of freedom or with phonons 
drastically reduces the kinetic energy \cite{b.a3}. Recently the CO is 
extensively 
 observed in real materials at high densities: hole ordering in 
 rare-earth pnictides like $\rm{Yb_4 As_3}$ 
\cite{b.a4}, charge ordering in the unconventional spin-Peierls material 
$\alpha'-\rm{NaV_2O_5}$
 \cite{b.a5} and in colossal magnetoresistance compounds, for example in 
 $\rm{R_{2-2x}A_{1+2x}Mn_2O_7  (R = La, Pr; A = Ca, Sr;}
 x \geq 0.5)$ \cite{b.a6}.\\

One of the simplest models of interacting electrons that allows for charge 
ordering is the extended Hubbard model (EHM). This model has been intensively 
studied both in low dimensions and in the limit of infinite dimension, 
usually at half or at quarter filling. A variety of techniques, such as 
Hartree-Fock approximation \cite{b.a7}, perturbation theory \cite{b.a8}, 
the dynamical mean field theory (DMFT) \cite{b.a9}, 
slave boson approach \cite{b.a10}, as well as numerical 
methods like Quantum Monte Carlo simulation \cite{b.a11} 
and Lanczos technique \cite{b.a12} 
have been employed. Obviously, each of these approaches is able to describe 
in a proper way only one of the relevant limits of the model and despite 
many publications devoted to the CO 
in solids the physical picture of the phenomenon is far from 
being clear.\\

Recently a melting of the charge ordered state on decreasing the
temperature has been found in the doped compounds
$\rm{Pr_{0.65}(Ca_{0.7}Sr_{0.3})_{0.35}MnO_3}$ \cite{b.a13} and 
$\rm{La_{2-2x}Sr_{1+2x}Mn_2O_7 }$ $(0.47 \leq x \leq 0.62)$ 
\cite{b.a14,b.a15}. 
A reentrant transition at quarter filling has been obtained theoretically 
using EHM both with electron-phonon interaction \cite{b.a16} and without 
electron-phonon interaction \cite{b.a9, b.a12}. 
The present paper is devoted to study of the 
boundary between the charge ordered and disordered phase for 
different regimes of the temperature $T$, the Coulomb interactions $U, V$ 
and the band filling $n$. A simple but physically meaningful 
approximation allowing to solve 
this problem is the  CPA. This self-consistent 
approximation is recognized as the best single-site approximation 
for the spectral properties of disordered systems. Originally, the 
alloy-analog approximation 
has been formulated as an approximation scheme for the Hubbard model 
\cite{b.a17}.
 To solve the alloy problem the CPA is used as a second step.
The CPA was also applied to intermediate valence and heavy fermion systems 
\cite{b.a18}. In the present work this approximation is used for the first 
time to treat the charge ordering in the EHM.\\   

\section {Model and formalism} 

 We consider the following Hamiltonian for the EHM:
\begin{eqnarray}
\label{1}
H &=& t {\sum_{<ij>\sigma }}(c_{i\sigma }^{+}c_{j\sigma }
 + c_{j\sigma }^{+}c_{i\sigma})  +\nonumber\\
&&U{\sum_{i}}n_{i\uparrow }n_{i\downarrow } +
V{\sum_{<ij>}}n_{i }n_{j}
\end{eqnarray}
where $c_{i \sigma} (c_{i \sigma}^+)$ annihilates (creates) an electron with 
spin $\sigma$ at site $i$,  $n_{i \sigma} = c_{i \sigma}^+c_{i \sigma}$ and 
$n_i = n_{i \uparrow} + n_{i \downarrow}. <ij>$ denotes nearest neighbors, 
$t$ is the hopping parameter, $U$ and $V$ are on-site and inter-site 
Coulomb repulsion, respectively. We divide the hypercubic lattice in two 
sublattices such that points on one sublattice have only points of the 
other sublattice as nearest neighbors. The sublattice is denoted by subindex 
A or B: $c_{i \sigma}= a_{i \sigma} (b_{i \sigma})$ if $i \in A \; (i \in B)$. 
Performing a mean-field decoupling of the $V$ term, we get
\begin{eqnarray}
\label{2}
H &=& {\sum_{i \in A, \sigma }}zVn_B a_{i\sigma }^{+}a_{i\sigma }  +
U{\sum_{i\in A}}n_{i\uparrow }n_{i\downarrow } +\nonumber\\
&&{\sum_{j \in B, \sigma }}zVn_A b_{j\sigma }^{+}b_{j\sigma }
+ U{\sum_{j\in B}}n_{j\uparrow }n_{j\downarrow }\\
&& + t {\sum_{<ij>\sigma }}(a_{i\sigma }^{+}b_{j\sigma }
 + b_{j\sigma }^{+}a_{i\sigma})- \frac{1}{2}zNVn_An_B \nonumber 
\end{eqnarray}
where $z$ is the number of nearest neighbors, $n_{A/B}$ is the averaged 
electron occupation number in the A/B-sublattice, $N$ is the number 
of sites in the lattice.\\
In the alloy-analog approach the many-body Hamiltonian (\ref{2}) is replaced 
by a one-particle Hamiltonian with disorder which is of the form
\begin{eqnarray}
\label{3}
H&=& {\sum_{i \in A, \sigma }}E_{A \sigma} a_{i\sigma }^{+}a_{i\sigma }  +
{\sum_{j \in B, \sigma }}E_{B \sigma} b_{j\sigma }^{+}b_{j\sigma }\nonumber\\  
&&+ t {\sum_{<ij>\sigma }}(a_{i\sigma }^{+}b_{j\sigma }
 + b_{j\sigma }^{+}a_{i\sigma}) - \frac{1}{2}zNVn_An_B  
\end{eqnarray}
where
\begin{equation}
\label{4}
E_{A/B, \sigma} = \cases{
zVn_{B/A}\quad \textrm{with probability}\quad  1 - n_{A/B, - \sigma}\cr 
zVn_{B/A}+ U \quad  \textrm{with probability}\quad  n_{A/B, - \sigma}\cr}
\end{equation}
In the following we assume spin-independent expectation values in 
(\ref{4}), i.e. 
we consider only nonmagnetic solution: $n_{\alpha \uparrow} = 
n_{\alpha \downarrow} = \frac{1}{2}n_{\alpha}; (\alpha = A, B)$. The 
Green function $G$ corresponding to the Hamiltonian (\ref{3}) has to be 
averaged 
over all possible configurations of the random potential which can be 
considered to be due to alloy constituents. The averaging cannot be performed 
exactly. To solve the alloy problem the CPA is used. The averaged 
Green function $\bar{G}$ is obtained from an effective Hamiltonian 
containing a self-energy  $\Sigma_{A/B}(\omega)$ for the A/B-sublattice:
\begin{eqnarray}
\label{5}
H_{eff}&=&\Sigma_A(\omega) {\sum_{i \in A, \sigma }} a_{i\sigma }^{+}
a_{i\sigma }  +
\Sigma_B(\omega){\sum_{j \in B, \sigma }} b_{j\sigma }^{+}b_{j\sigma }
\nonumber\\  
&&+ t {\sum_{<ij>\sigma }}(a_{i\sigma }^{+}b_{j\sigma }
 + b_{j\sigma }^{+}a_{i\sigma}) 
\end{eqnarray}
In momentum space the averaged Green functions
$\bar{G}_{A/B}(\vec{k},\omega)$ for the A/B-sublattice are of 
the form  
\begin{eqnarray}
\bar{G}_A(\vec{k}, \omega)&=& \left(\omega -\Sigma_A(\omega) - \frac{t_{
\vec{k}}^2}{\omega - \Sigma_B(\omega)} \right)^{-1}\\
\bar{G}_B(\vec{k}, \omega)&=& \left(\omega -\Sigma_B(\omega) - \frac{t_{
\vec{k}}^2 }{\omega - \Sigma_A(\omega)} \right)^{-1}
\end{eqnarray}
where $t_{\vec{k}}$ is the Fourier transform of the hopping matrix element 
(the wavevector $\vec{k}$ is for sublattice A (or B)), $\Sigma_{A/B}(
\omega)$ are to be determined latter.\\
The density of states (DOS) for free electrons with the band dispersion $t_{
\vec{k}}$ is replaced by the semi-elliptical $\,\rho_0(\omega) = \frac{2}
{\pi W^2}\sqrt{W^2-\omega^2}$ (we set $W$ = 1 as the unit for the 
energy scale). Note that this model DOS is often used as an additional 
 approximation in combination with the CPA. It was noted in Ref.\cite{b.a19} 
that for the Bethe lattice with $3 \leq z \leq 6$ this approximation
is a good one, at least in a qualitative sense.  The averaged Green functions 
$\bar{G}_{A/B}(\omega)$ for the A/B-sublattice then take the form

\begin{eqnarray}
\label{8}
\bar{G}_A(\omega) &=& \frac{1}{N}\sum_{\vec{k}}
\bar{G}_A(\vec{k}, \omega)
= \frac{2}{W^2} \biggl(\omega -\Sigma_B(\omega)\nonumber\\ 
&&- \left[ (\omega - \Sigma_B (\omega))^2 -\frac{\omega - \Sigma_B (\omega)}
{\omega - \Sigma_A(\omega)}W^2 \right]^{1/2}\biggr)
\end{eqnarray}

And $\bar{G}_B(\omega)$ is obtained by replacing $A \leftrightarrow B$.
A scattering matrix $T$ is introduced for each configuration via
\begin{equation}
\label{9}
G = \bar{G} + \bar{G}T\bar{G}
\end{equation}
The CPA demands that the scattering matrix vanishes on average: $\bar{T} 
= 0$. This yields expression for $\Sigma_{A/B}(\omega)$ of the form
\begin{eqnarray}
\label{10}
\Sigma_A(\omega)&=& \bar{E}_A - (zVn_B - \Sigma_A(\omega))\cdot\nonumber\\ 
&&\bar{G}_A (\omega)(zVn_B + U - \Sigma_A(\omega))
\end{eqnarray}
where $\bar{E}_{A} = zVn_{B} +\frac{1}{2}Un_{A}$. Again  $\Sigma_B(\omega)$ 
and $\bar{E}_B$ are obtained by replacing $A \leftrightarrow B$. \\
For arbitrary size of electron density $n$ we make the following ansatz:
$$n_{A/B}=n \pm x;\quad \bar{G}_{A/B}(\omega) = G(\pm x, \omega)$$ 
Eliminating 
$\Sigma_A(\omega), \Sigma_B(\omega)$ from (\ref{8}) and (\ref{10}) 
leads to an 
equation for $G(\pm x, \omega)$:
\begin{eqnarray}
\label{11}
&&\omega - \frac{G(-x,\omega)}{4}- \frac{1}{G(x,\omega)}=zV(n-x) + 
\frac{U}{2}(n+x) -\nonumber\\ 
&&\bigg[ zV(n-x) - \omega 
+ \frac{G(-x,\omega)}{4} + \frac{1}{G(x, \omega)}\bigg]\cdot\\
&&\bigg[ zV(n-x) + U - \omega + \frac{G(-x,\omega)}{4} + \frac{1}{G(x, \omega)}
\bigg] G(x,\omega)\nonumber
\end{eqnarray}
Setting $x = 0$ in  eq.(\ref{11}) and shifting the one-electron energy by 
$zVn + \frac{U}{2}$ we reproduce the CPA equation for the Green function 
obtained by Velicky et al in Ref.\cite{b.a20}.\\
If $\mu$ is the chemical potential of electrons, then at the temperature 
$T$ one has
\begin{eqnarray}
\label{12}
n_{\alpha} &=& \frac{2T}{N}{\sum_{\vec{k},n}}\bar{G}_{\alpha}
(\vec{k}, i \omega_n)\nonumber\\ 
&=&- \frac{2}{\pi}\int_{-\infty}^{+ \infty}
 d \omega f(\omega)\Im G(n_{\alpha}-n, \omega)
\end{eqnarray}
where $\omega_n = (2n+1)\pi T$ is the Matsubara frequencies and 
$f(\omega) = (1 + \exp (\omega-\mu)/T)^{-1}$ is the Fermi function.\\
The pair of equations (\ref{12}) must now be solved with the constraint 
$n_A + n_B = 2n$ for $n_A, n_B$ and $\mu$. For small enough $V$ the solution 
of (\ref{12}) is the charge disordered state with $n_A= n_B$. But if $V$ is 
sufficiently large it may also be possible to find the CO solution for 
which $n_A \neq n_B$. One finds that the condition for the onset of 
charge ordering is equivalent to  $n_A = n_B = n$ being a double solution 
of (\ref{12}). This condition is expressed as
\begin{eqnarray}
\label{13}
n&=& - \frac{2}{\pi}\int_{-\infty}^{+ \infty}
d \omega f(\omega)\Im G(0, \omega)\\
\label{14}
1 &=& - \frac{2}{\pi}\int_{-\infty}^{+ \infty}
d \omega f(\omega)\Im G'(0, \omega)
\end{eqnarray}
where $G'(0,\omega) = \frac{\partial G(x, \omega)}{\partial x} \mid_{x=0}$ 
and $G(0, \omega)$ is a solution of (\ref{11}) when $x=0$. 
The latter is a cubic 
equation for $G(0,\omega)$ and the correct root must be identified  from 
the physical condition to yield a non-negative density of states. 
It is also easy to 
obtain $G'(0, \omega)$ from eq.~(\ref{11}). So, for fixed temperature 
$T$, on-site 
Coulomb repultion $U$ and band filling $n$ we have the closed system of 
equations (\ref{13})-(\ref{14}) for the critical value $V$ and the 
chemical potential $\mu$ within the framework of the CPA.\\ 

 \begin{figure}
 \begin{center}
 \psfig{figure=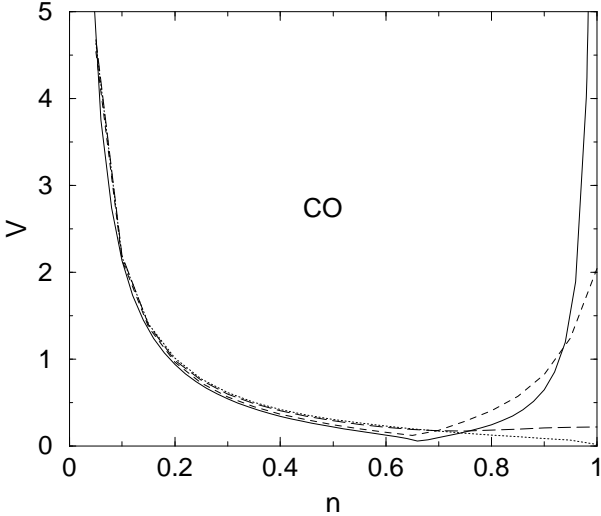,width=8cm}
 \end{center}
 \caption{$(V$-$n)$-phase diagram for the 2D extended Hubbard model 
 $(W = 1,\: T = 0)$ for different values of $U: \: U = 0,\,0.5, \, 
 1.5 \,$ and $ U = \infty \,$ corresponding to the dotted, long-dashed, 
 dashed and solid lines, respectively.}
 \label{Fig1}
 \end{figure}
 \begin{figure}
 \begin{center}
 \psfig{figure=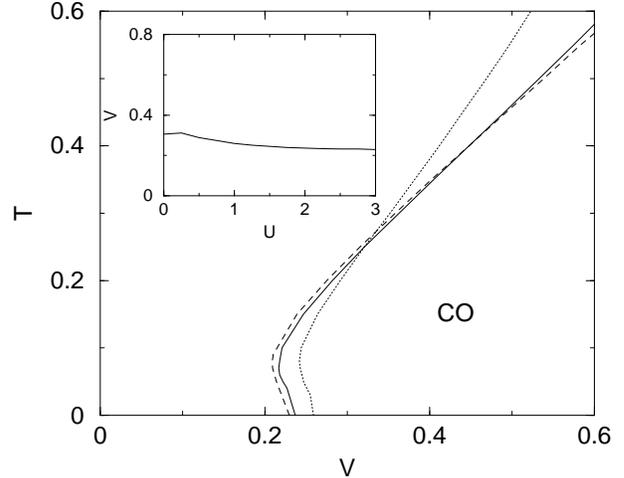,width=8cm}
 \end{center}
 \caption{$(T$-$V)$-phase diagram at quarter filling for several values 
 of $U: \: U = 1,\, 2\,$ and 3 corresponding the dotted, solid and dashed 
 lines, respectively. The inset shows $(U$-$V)$-phase diagram for $n = 1/2$ 
 and $T = 0$.}
 \label{Fig2}
 \end{figure}

\section {Numerical results and discussion}  

 We have solved numerically eqs.~(\ref{13})-(\ref{14}), the results 
may be summarized as follows. In Fig.~\ref{Fig1} phase 
diagram as a function of $n$ 
and $V$ for different values of $U$ for the two-dimensional square lattice 
at zero temperature. The half-bandwith $W$ was taken as unit of energy 
(for the square lattice $z=4$ and $W= 4t$). Due to the electron-hole 
symmetry we consider only $0 \leq n \leq 1$. From Fig.~\ref{Fig1} one
can see that 
in the two regions $(n \ll 1,\, n \leq 1)$ the influence of $U$ 
on the boundary between the charge ordered and the homogenous disordered 
phase is different: 
away from half filling $(n < n^* \approx 0.67)$ the on-site interaction 
$U$ has little effect on the critical value $V_c$, while for $n^* < n 
\leq 1$ the critical $V_c$ strongly depends on $U$. In Ref.\cite{b.a15} 
Dho et al 
found that CO exists over a broad doping range $(0.44 \le x \le 0.8)$ in 
La$_{2-2x}$Sr$_{1+2x}$Mn$_2$O$_7$. It is worthwhile to note that for 
large $U$ in the same filling region the CPA values of $V_c$ are close 
the minimum value ($V_c$ has the minimum at $n^* \approx 0.67$). Although 
the mechanism of the charge ordering in the layered manganites is more 
complex than one induced by a nearest neighbor Coulomb repulsion, from 
the above result we may speculate that the CPA is able to describe the 
CO boundary in this compound. In addition, the advantage of the CPA is 
that by using this simple approach one can easily obtain phase diagrams 
in the $(V$-$n)$-plane for arbitrary $U$ as well as in the $(U$-$V)$-plane 
at arbitrary $n$. As an illustration of our approach in the following we 
consider CO transition on the square lattice at quarter filling $(n = 
1/2)$. The inset in Fig.~\ref{Fig2} shows the $(U$-$V)$-phase diagram at zero 
temperature. We 
compare CPA result to the one obtained by other methods. At $U = \infty$ 
the critical value $V_c = 0.195W$ and $V_c = 0.172W$ obtained in 
Ref.\cite{b.a10} 
by the slave boson approach with a constant and the actual DOS $\rho_0 (
\omega)$ respectively, is in a good agreement with our result $V_c = 0.218W$. 
At $U= 2W \; V_c = 0.66W$ was obtained in Ref.\cite{b.a9} for the EHM in 
infinite dimension by the numerical renormalization group (NRG) method, 
 while the CPA result for our 2D lattice is $V_c = 0.237W$.

 \begin{figure}
 \begin{center}
 \psfig{figure=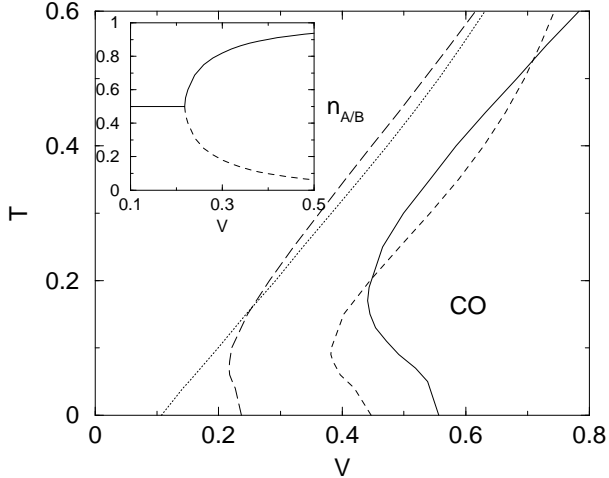,width=8cm}
 \end{center}
 \caption{$(T$-$V)$-phase diagram for $U = 2$ at various values of $n:\: 
 n = 0.3,\,0.5, \, 0.65 \,$ and 0.8 corresponding to the 
 solid, long-dashed, dotted and dashed lines, respectively. The inset shows 
 the lattice occupancies $n_A$ (solid line) and $n_B$ (dashed line) as 
 a function of $V$ for $U = \infty$ and $T = 0$.}
 \label{Fig3}
 \end{figure}
 \begin{figure}
 \begin{center}
 \psfig{figure=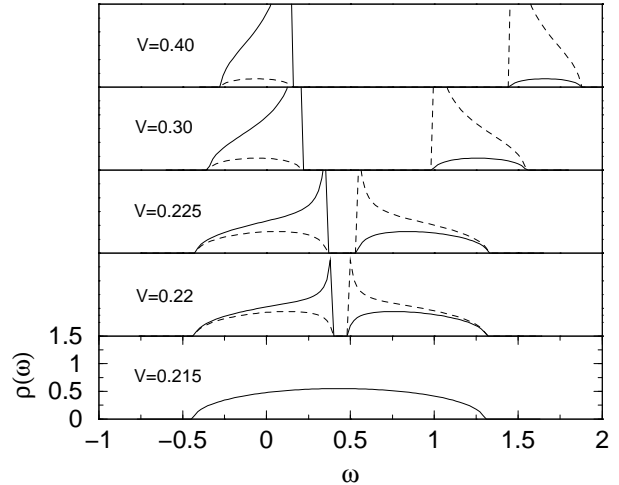,width=8cm}
 \end{center}
 \caption{CPA results at quarter filling for the A and B spectral functions 
 $\rho(\omega)$ (solid and dashed lines, respectively) for 
 $U = \infty,\, T = 0$ and several values of $V$. Below the critical 
 $V_c \approx 0.218$, the spectral functions of both sublattices are equal.}
 \label{Fig4}
 \end{figure}

In Ref.\cite{b.a10} McKenzie et al argue that in the large $U$ limit 
at quarter filling the charge ordered phase is destroyed below a critical 
non-zero value $V_c$, of the order of $t$, while we can now show in 
the inset of Fig.~\ref{Fig2} 
that the critical value $V_c$ is almost independent of $U$, therefore $V_c$ 
is of the order of $t$ for all values of $U$. Fig.~\ref{Fig2} shows the 
$(T$-$V)$-phase diagram for different values of $U$. 
For each value of $U$ reentrant 
behavior as function of temperature is seen for some region of $V$. For $V$ 
in this range, the ground state is homogeneous but a charge ordered phase 
exists at intermediate temperature. Note that reentrant transition 
apparently is not obtained by Hartree-Fock approximation in the EHM without 
electron-phonon interaction. Fig.~\ref{Fig3} shows the $(T$-$V)$-phase diagram 
for $U = 2$ at 
different band fillings $n$. The reentrant behavior is seen clearly for 
the values of $n$ where $V_c(T=0)$ is larger (e.g. $\, n=0.3,\, n=0.8$), 
while for values of $n$ where $V_c(T=0)$ is small (e.g. $\, n=0.65$), 
the reentrant behavior is not seen.\\
In the inset in Fig.~\ref{Fig3} we present the CPA result for $V$ dependence 
of the sublattice 
occupancies $n_A$ and $n_B$ and hence the charge order parameter $(n_A - 
n_B)$ in the strong correlated limit $U \rightarrow \infty$ at zero 
temperature. The transition is clearly continuous in contrast to the 
result for $U = 2,\, T = 0$ in Ref. \cite{b.a9} where the NRG method gives 
a first order phase transition. Note that in the strong correlated limit 
$U \rightarrow \infty$ from eq.~(\ref{11}) one can find an analytic 
expression 
for $G(x, \omega)$. Then eqs.~(\ref{12}) have to solved self-consistently 
to find 
$n_A, n_B$ and $\mu$. For a given set of these parameters the total energy 
of the system for various states can be calculated. The state with the 
lowest energy is the true ground state and determines the spectrum. The 
CPA result for A and B spectral functions for $U = \infty,\, T = 0$ are 
shown in Fig.~\ref{Fig4}. When $V < V_c \approx 0.218$ the A and B spectral 
functions are identical and they are independent of $V$ by shifting all 
the one-electron energy levels and the chemical potential by $2V$. For 
$V > V_c$ the A and B spectral functions change and each spectrum splits 
in the upper and lower subbands. On increasing $V$ the weight of the 
lower subband in the A-spectrum increases, while the one in the B-spectrum 
decreases; the subbands in the spectrum become narrower due to the reduced 
hopping of electrons in the charge ordered phase. A charge ordering gap 
opens and it is given by $\Delta = 2 \sqrt{16 V^2x^2 + \gamma}\,$, where 
$\,\gamma = \frac{1}{2}[(1-\frac{n}{2}) - \sqrt{(1 - \frac{n_A}{2})(1 - 
\frac{n_B}{2})}]$. In the limit of large $V$, perfect charge order evolves: 
$n_A \rightarrow 1,\, n_B \rightarrow 0$. Therefore in this limit 
$x \rightarrow 1/2$ and the gap between two peaks $\Delta$ is given 
in CPA by $4V$, as compared to $2V$ obtained by the DMFT in Ref.\cite{b.a9}.\\

\section {Conclusions} 
 In this paper we have applied the CPA to study charge ordering in the 
 extended Hubbard model. Within this approximation one can obtain the 
critical value $V_c$ as a function of temperatute $T$, on-site Coulomb 
repulsion $U$ and band filling $n$. To examine the CPA results we consider 
the charge ordering transition on the 2D square lattice at quarter filling 
using a semi-elliptical DOS. It was shown that for all values of $U$ the 
charge ordered phase is destroyed below a critical non-zero value $V_c$ 
of the order of $t$. Like previous results in Ref. \cite{b.a9,b.a12}, we find 
a parameter region where the model shows reentrant behavior. The reentrant 
transition is also observed at other band filling, as it was experimentally 
found in the layered manganites in Ref. \cite{b.a15}. In the strong 
correlation limit $U \rightarrow \infty$ at zero temperature the CPA 
gives a continuous transition.\\
Now the CPA is known to give good results of one-particle properties 
in a wide range of systems.  To study the CO boundary 
phase in the EHM the CPA has the advantage over DMFT of being analytically 
simple, and over Hartree-Fock approximation (small $U$) and slave boson 
approach $(U \rightarrow \infty)$ of being able to describe the whole range 
for the on-site interaction $U$. Of course in common with the alloy CPA 
the imaginary part of the self-energy does not vanish at the Fermi level 
at $T = 0$, so we do not obtain a true Fermi liquid and we only expect our CPA to study CO in the EHM well at finite temperature.\\
The calculation presented here can also be applied to the lattice of 
higher dimensions, or to the EHM in the presence of a weak magnetic field. 
To include magnetic phases and cluster effects one has to go beyond 
the usual alloy CPA. This is left for our future work.\\[0.2cm]
{\bf Acknowledgments}\\[0.2cm]
The authors would like to thank Prof. 
P. Fulde for useful discussions and Dr. B. Schmidt 
for the help in 
numerical calculation. This work has been done during a visit of A.T. 
Hoang to the MPI PKS, Dresden, whose hospitality and support are gratefully 
acknowledged. \\[0.5 cm]



\newpage

\begin{references}
\bibitem{b.a1}  Wigner E 
1938 {\it Trans. Faraday Soc.} {\bf 34} {678}.
\bibitem{b.a2}  Andrei E Y {\it  et al.} 
1988 {\it Phys. Rev. Lett.} {\bf 60} {2765}.
\bibitem{b.a3} Fulde P 1997  {\it Ann. Phys.} {\bf 6} {178}.
\bibitem{b.a4} Ochiai A, Suzuki T and Kasuya T 
1990 {\it J. Phys. Soc. Jpn.} {\bf 59} {4129}.
\bibitem{b.a5} Ohama T  {\it et al.} 
1999 {\it Phys. Rev. } {\bf B 59} {3299}.
\bibitem{b.a6} Chen C H and Cheong S W 
1996 {\it Phys. Rev. Lett.} {\bf 76} {4042}.
\bibitem{b.a7} Seo H and Fukuyama H 
1998 {\it J. Phys. Soc. Jpn.} {\bf 67} {2602}.
\bibitem{b.a8} van Dongen P G J 
1994 {\it Phys. Rev. } {\bf B 50} {14016}.
\bibitem{b.a9} Pietig R, Bulla R and Blawid S 
1999 {\it Phys. Rev. Lett.}{\bf 82} {4046}.
\bibitem{b.a10} McKenzie R H  {\it et al.} 
2001 {\it Phys. Rev. } {\bf B 64} {085109}.
\bibitem{b.a11} Hirsch J E 
1984 {\it Phys. Rev. Lett.} {\bf 53} {2327}.
\bibitem{b.a12} Hellberg C S
2001 {\it J. Appl. Phys.} {\bf 89} {6627}.
\bibitem{b.a13}  Tomioka Y  {\it et al.} 
1997 {\it J. Phys. Soc. Jpn.} {\bf 66} {302}.
\bibitem{b.a14} Chatterji T {\it et al.} 
2000 {\it Phys. Rev. } {\bf B 61} {570}.
\bibitem{b.a15}  Dho J { \it et al.} 
2001 {\it J. Phys.: Cond. Matt.} {\bf 13} {3655}.
\bibitem{b.a16}  Yuan Q and Thalmeier P 
1999 {\it Phys. Rev. Lett.} {\bf 83} {3502}.
\bibitem{b.a17} Hubbard J
1964 {\it Pro. R. Soc. London} {\bf A 281} {401}.
\bibitem{b.a18} Czycholl G
1986 {\it Phys. Reports} {\bf 143} {277}.
\bibitem{b.a19} Vlaming R and Vollhardt D 
1992 {\it Phys. Rev. } {\bf B 45} {4637}.
\bibitem{b.a20} Velicky B, Kirkpatrick S and Ehrenreich H 
1968 {\it Phys. Rev.} {\bf 175} {747}.
\end{references}
\end{document}